# WSPEC: A Waveguide Filter Bank Spectrometer

George Che[1*], Sean A. Bryan[1], Matthew Underhill[1], Philip Mauskopf[1], Christopher Groppi[1],
Glenn Jones[2], Bradley R. Johnson[2], Heather McCarrick[2], Daniel Flanigan[2], and Peter Day[3]

[1]*Arizona State University, Tempe, AZ 85287, USA*
[2]*Columbia University, New York, NY 10027, USA*
[3] *Jet Propulsion Laboratory, Pasadena, CA 91109, USA*
*Contact: gche2@asu.edu*

*Abstract*— We have designed, fabricated, and measured a 5-channel prototype spectrometer pixel operating in the WR10 band to demonstrate a novel moderate-resolution ($R = f/\Delta f \sim 100$), multi-pixel, broadband, spectrometer concept for mm and submm-wave astronomy. Our design implements a transmission line filter bank using waveguide resonant cavities as a series of narrow-band filters, each coupled to an aluminum kinetic inductance detector (KID). This technology has the potential to perform the next generation of spectroscopic observations needed to drastically improve our understanding of the epoch of reionization (EoR), star formation, and large-scale structure of the universe. We present our design concept, results from measurements on our prototype device, and the latest progress on our efforts to develop a 4-pixel demonstrator instrument operating in the 130-250 GHz band.

## I. Introduction

Technological advancements in imaging and spectroscopy in the mm and submm-wave regimes have revolutionized the fields of observational cosmology and extragalactic astronomy. ALMA, a product of these advancements, is currently performing spectroscopic measurements at resolutions and sensitivities much higher than those ever previously attainable. While ALMA is a superb tool for performing high-resolution imaging and spectroscopy on individual sources, it would be prohibitively time-consuming to use it for wide-band spectral surveys over large areas of the sky. Requiring only moderate spectral resolution ($R \sim 50 - 200$), such surveys are vitally important to the challenging next steps in mm-wave imaging and spectroscopy, which aim to characterize the large-scale structure and star formation history of the universe using CO and CII intensity mapping and perform high angular resolution observations of the hot gas in galaxy clusters using the SZ effect.

Tackling these challenges requires a multi-pixel, broadband spectrometer comprised of compact spectrometer arrays coupled to large arrays of highly-multiplexable detectors. The current state of the art in mm-wave spectroscopy is Z-Spec [1], a single-pixel grating-type spectrometer that achieves $R \sim 300$. There are also substantial ongoing efforts to develop ultra-compact on-chip spectrometers (e.g. SuperSpec [2] and DESHIMA [3]) based on lithographically-patterned superconducting filter banks coupled to large arrays of KIDs. We are developing a scalable multi-pixel waveguide spectrometer (WSPEC) that implements filter banks using rectangular waveguide resonant cavities instead of lithography for horn-coupled imaging spectroscopy using KIDs. The spectrometer pixels, which can be warm and cold-tested independently from the detector arrays, are fabricated with standard precision-machining tools. WSPEC is a highly complementary technology to on-chip designs in several ways: 1) Our WSPEC demonstrator instrument targets the relatively unexplored 130-250 GHz band, which is suitable for CO line emission and kinetic SZ studies, 2) WSPEC is designed for lower spectral resolution than the superconducting spectrometers, and 3) WSPEC may be used as a room-temperature backend for cryogenic amplifiers, removing the need for down-converting mixers.

## II. Design

### A. Waveguide Filter Bank Concept

The design of a single waveguide spectrometer pixel is illustrated in the top panel of Fig. 1, which shows an HFSS drawing of our 5-channel prototype filter bank. A feed horn couples light from the sky into the main waveguide. Each channel connects to the main waveguide through an evanescent coupling section into a $\lambda/2$ resonant cavity, the electrical length of which defines the center frequency of the channel. An identical coupling section on the other side of the resonator connects to another section of waveguide, which is terminated by an aluminum KID. Since the cutoff frequency of the coupling sections is approximately 1.5 times the channel's center frequency, these sections are seen as capacitive loads. On-resonance, the cavity becomes an inductive load that tunes out the capacitive sections allowing a narrow band of light centered on the resonant frequency to propagate through to that channel's KID. Off-resonance, no impedance cancelation occurs, so no light passes through. Therefore, each channel is a narrow-band frequency filter.

### B. WR10 Prototype Pixel

We have successfully designed, fabricated, and tested a 5-channel prototype filter bank for the WR10 band. The reason for choosing this band is two-fold: 1) we own a WR10 VNA extender and 2) the relatively large dimensions of WR10 waveguide are suitable for a first fabrication. We designed the prototype to have five $R\sim100$ channels centered at the following frequencies: $f_c = 80$ GHz, three closely-spaced channels near 90 GHz, and $f_c = 105$ GHz. The 80 GHz and 105 GHz channels span the WR10 band and the middle three channels are spaced in frequency according to the geometric progression described in [4]. We chose $3\lambda/4$ physical spacing between adjacent channels, where $\lambda$ is the average wavelength of the two channels, because this is the smallest physically realizable odd integer multiple of $\lambda/4$. Individual channels were optimized in HFSS to obtain the appropriate dimensions for each channel before fabrication.

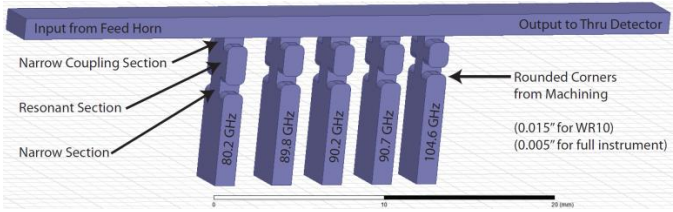

Fig. 1 HFSS drawing of our 5-channel WR10 prototype spectrometer pixel.

Just as planned for the full demonstrator instrument, we employed E-plane split-block construction using conventional alignment pins, as shown in the top-left panel of Fig. 3. The prototype device was machined from aluminum using a 5 μm-tolerance micromilling machine at ASU. The 1 μm tolerance required for the higher-frequency full instrument is consistently achieved on another machine in our lab at ASU.

III. SIMULATION AND MEASUREMENT RESULTS

After optimizing the dimensions for each channel, we performed a final simulation of the entire 5-channel prototype device in HFSS. The full structure was small enough for the simulation to finish in a single day and the results are shown in the top panel of Fig. 2. We see that the full simulation closely matches the designed center frequencies and resolving power for each channel.

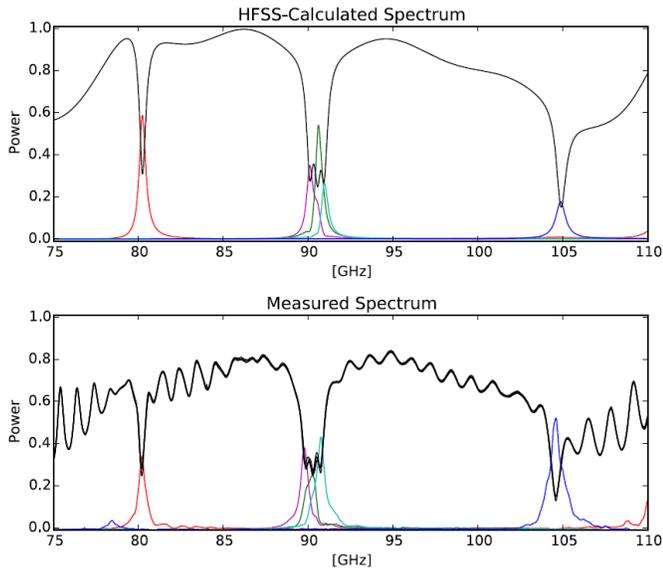

Fig. 2 Simulated (top) and measured (bottom) power absorbed by each channel and thru detector of our 5-channel WR10 prototype device.

As shown in Fig. 3, we measured the absorption efficiency of each channel of our device by using standard gain horns from Quinstar to terminate all but one channel to absorbing AN-72 foam and connecting diode detectors to the remaining channel and thru port. Our VNA was in unavailable when we performed this measurement, so we generated mm waves using the source component of our WR10 VNA extender driven by a signal generator. The efficiency of each channel was obtained by sweeping through the WR10 band at the source and recording the signal measured on the diode detector connected to each channel in succession and terminating all other channels with horns. The results are shown in the bottom panel of Fig. 2.

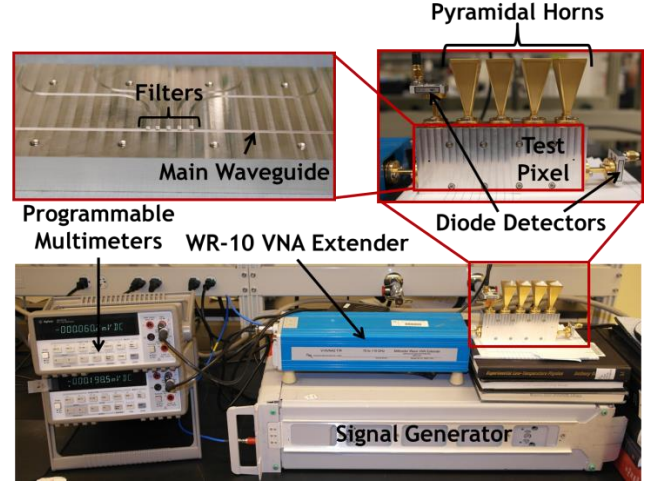

Fig. 3 Laboratory setup for measuring our WR10 device (bottom), featuring a close-ups of the device connected to horns and detectors (top-right) and half of the device, illustrating E-plane split-block construction (top-left).

Measured center frequencies agree with the simulation to within 0.5% and resolving powers to within 30%. The standing wave pattern observed in the thru detector spectrum is possibly accounted for by imperfect terminations in the horns or detectors and/or a mismatch inside the VNA extender. We will perform additional measurements using our VNA in the near future, but overall, our measurement results agree with HFSS simulations, confirming that our device at least nominally works according to design and suggesting HFSS as an appropriate tool to design our full demonstrator instrument.

IV. 4-PIXEL DEMONSTRATOR SPECTROMETER

An important next step is demonstrating a small array of waveguide spectrometer pixels coupled to arrays of KIDs targeting a scientifically interesting frequency band. We have decided to construct a 4-pixel array of filter banks targeting the 130-250 GHz band, which is optimal for studying CO line emission and the kinetic SZ effect.

In order to Nyquist sample the entire 130-250 GHz band, we need 108 channels. The entire band is too wide for a single-mode rectangular waveguide, so we will split the band into a lower band A below the 183 GHz atmospheric line and upper band B above the line. Two independent horns will feed 54-channel filter banks covering bands A and B and these two filter banks collectively comprise a single spatial pixel. As illustrated in Figure 4, which shows a Solidworks drawing of our 4-pixel design, a linear array of spatial pixels is formed in the Y-direction, with all spectral channels feeding a single card of KIDs. These linear arrays are then stacked in the Z-direction to form a filled 2-dimensional focal plane array of filter bank spectrometer pixels. Using $3\lambda/4$ physical spacing between adjacent channels, the band A and B filter banks excluding horns, are only 96 mm and 68 mm long, respectively. Even a 100-pixel instrument, an eventual goal for this technology, would still be relatively compact.

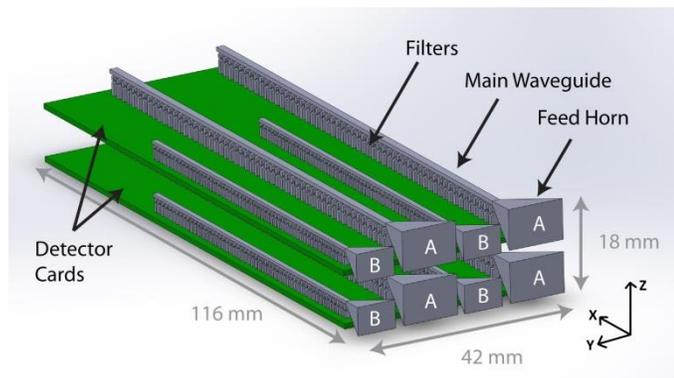

Fig. 4 Solidworks drawing of our horn-coupled 4-pixel focal plane array of filter banks.

Drawing and simulating an entire 54-channel filter bank is prohibitively memory-intensive and time-consuming to do in a single HFSS run. Therefore, we developed an equivalent method, which entails using HFSS to compute the scattering matrix of each individual channel and then cascading these matrices together with the scikit-rf package in python. Using this method, we can reproduce all the details of a full HFSS simulation down to the -60 dB level and simulate an entire 54-channel filter bank in only 3 hours of computer time.

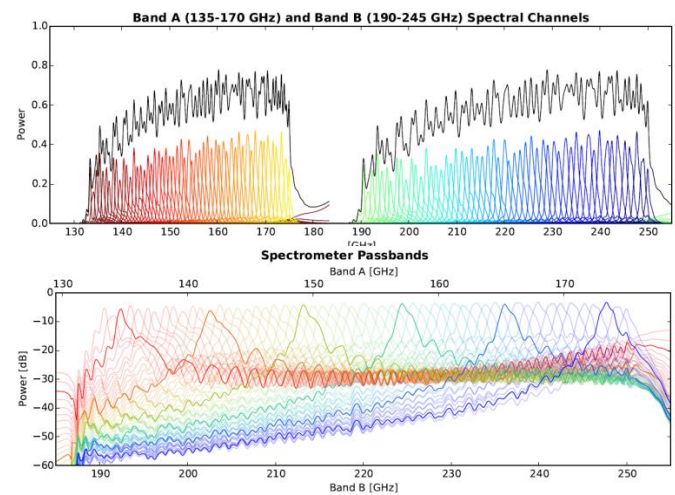

Fig. 5 HFSS-simulated passbands in linear units (top) and dB (bottom) for band A, which covers 135-170 GHz, and band B, which covers 190-245 GHz. The black curve in the top panel represents the sum total of all the passbands.

We simulated both band A and band B filter banks and the results are shown in both linear units and dB in Fig. 5. The absorption efficiency of the channels ranges from 0.25 to 0.4 and the out-of-band coupling is at the -20 to -30 dB level.

V. DISCUSSION AND FUTURE WORK

Before proceeding further, we must consider waveguide loss, which can severely limit the performance of mm-wave devices. The good agreement between lossless-conductor HFSS simulations and measurements of our prototype implies that conductor loss is not limiting the performance of our WR10 device at room temperature. However, scaling the design to higher frequencies could make conductor loss a more significant factor. Operating the device at cryogenic temperatures and switching from aluminum to a different material such as gold-plated OFHC copper will reduce loss. We will perform more detailed HFSS simulations to decide on the material to use for the full instrument.

In the short term, we will perform additional measurements on our prototype pixel using our VNA to better characterize the device. We will then fabricate and test a similar pixel that is scaled to band A in the 130-250 GHz band and investigate coupling its channels to a linear array of KIDs. The KIDs will be fabricated at ASU's nanofabrication center. When coupling KIDs to channels is well-understood, we will begin constructing the full 54-channel filter bank pixels, which will be tested individually using mm-wave sources at ambient and cryogenic temperatures. The long-term goal is to construct the full horn-coupled array of spectrometer pixels, mount it in an existing 230 mK cryogenic system at ASU and deploy the instrument on a telescope within two years.

VI. CONCLUSIONS

We presented a novel mm-wave spectrometer technology that has the potential to markedly improve our understanding of star formation and large-scale structure of the universe. Measurements of our WR10 prototype device have yielded promising results and we are currently developing a 4-pixel demonstrator instrument that operates in the 130-250 GHz band. In the future, we may investigate the possibility of incorporating an orthomode transducer and wide band frequency diplexer to allow dual polarization measurements over the entire frequency band for each pixel.